\documentclass[aps,prl,twocolumn,groupedaddress,showpacs]{revtex4}
\usepackage{graphicx}

\begin{document}

\newcommand{\etal}{\emph{et al.}}
\newcommand{\Li}{{}^{6}\textrm{Li}}
\newcommand{\K}{{}^{40}\textrm{K}}
\newcommand{\Rb}{{}^{87}\textrm{Rb}}
\newcommand{\kB}{k_{\mathrm{B}}}
\newcommand{\Rbmp}{\textrm{Rb}\left|2,2\right\rangle}
\newcommand{\RbTwoOne}{\textrm{Rb}\left|2,1\right\rangle}
\newcommand{\RbTwoZero}{\textrm{Rb}\left|2,0\right\rangle}
\newcommand{\RbOneZero}{\textrm{Rb}\left|1,0\right\rangle}
\newcommand{\RbOneOne}{\textrm{Rb}\left|1,1\right\rangle}
\newcommand{\RbOneMOne}{\textrm{Rb}\left|1,-1\right\rangle}
\newcommand{\Kmp}{\textrm{K}\left|9/2,9/2\right\rangle}
\newcommand{\Limp}{\textrm{Li}\left|3/2,3/2\right\rangle}

\title{Quantum degenerate two-species Fermi-Fermi mixture coexisting with a Bose-Einstein condensate}
\author{M.~Taglieber}\email[Electronic address: ]{matthias.taglieber@mpq.mpg.de}
\author{A.-C.~Voigt}
\author{T.~Aoki}
\author{T.W.~H\"ansch}
\author{K.~Dieckmann}
\affiliation{Department f\"ur Physik der
Ludwig-Maximilians-Universit\"at, Schellingstr.\,4, 80799 Munich,
Germany and Max-Planck-Institut f\"ur Quantenoptik, Hans-Kopfermann-Str.1, 85748 Garching,
Germany}

\date{\today}

\begin{abstract}
We report on the generation of a quantum degenerate Fermi-Fermi mixture of two different atomic species. The quantum degenerate mixture is realized employing sympathetic cooling of fermionic $^{6}$Li and $^{40}$K gases by an evaporatively cooled bosonic ${}^{87}\textrm{Rb}$ gas. We describe the combination of trapping and cooling methods that proved crucial to successfully cool the mixture. In particular, we study the last part of the cooling process and show that the efficiency of sympathetic cooling of the $^{6}$Li gas by $^{87}$Rb is increased by the presence of $^{40}$K through catalytic cooling. Due to the differing physical properties of the two components, the quantum degenerate $^{6}$Li-$^{40}$K Fermi-Fermi mixture is an excellent candidate for a stable, heteronuclear system allowing to study several so far unexplored types of quantum matter.
\end{abstract}


\maketitle

\begin{figure}
 \centering
 \includegraphics[width=7cm]{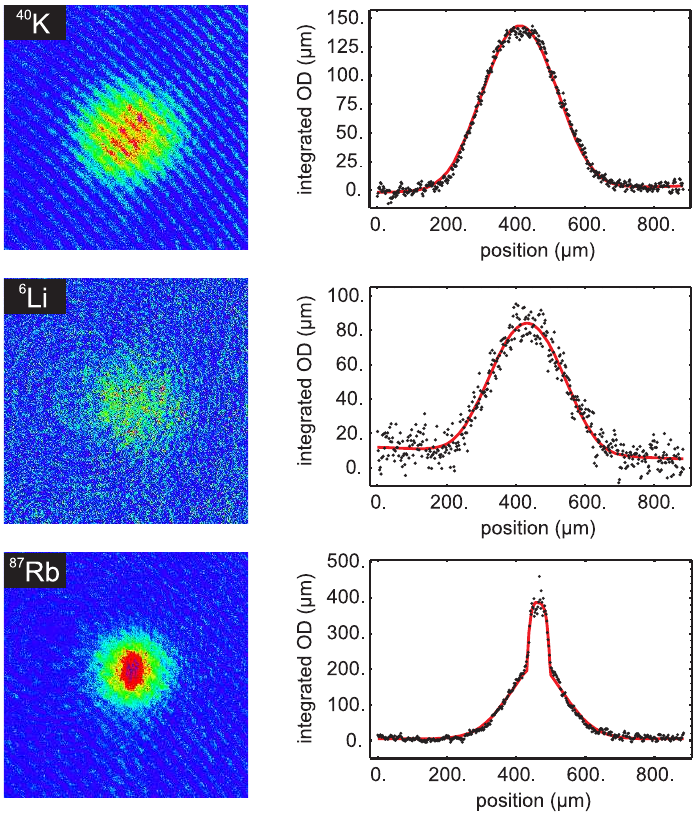}
 \caption{\label{fig:Clouds} Time of flight absorption images of the quantum degenerate Fermi-Fermi-Bose mixture. Shown are the optical column densities (left) and projections of the optical column densities (right) along with corresponding fits (lines). The projections are taken along the symmetry axis of the trap (x-axis) for the fermionic species and along the radial axis (y-axis) for rubidium. Expansion times are \mbox{15\,ms}, \mbox{4\,ms} and \mbox{20\,ms} for potassium, lithium and rubidium, respectively.}
\end{figure}

During the past decade, atomic physics has seen a revolution with the realization of new quantum matter like Bose-Einstein condensates (BEC) \cite{BEC} and the Mott insulator state \cite{Greiner2002}. More recently, the focus of interest shifted to ultracold fermionic quantum gases \cite{Fermions}, which allowed to study the crossover regime between a molecular Bose-Einstein condensate (BEC) and a Bardeen-Cooper-Schrieffer (BCS) like gas of paired fermions \cite{BecBcs}. Current research aims at simulating correlated many-body quantum systems with ultracold gases. A particularly intriguing goal is the realization of a fermionic quantum gas with two \emph{different} atomic species, which is a well controllable system and is predicted to be stable \cite{FermiFermi}. Due to the mass difference, it offers a variety of analogies to other many-body systems, in particular to a spatially inhomogeneous superfluid phase predicted to occur in certain types of high temperature superconductors \cite{ManyBody}. Further, a transition to a cristalline phase in the bulk gas \cite{Petrov2007} and the possibility to simulate baryonic phases of QCD \cite{QCD} have been theoretically proposed. Moreover, the mixture bears the prospect to create heteronuclear ground state molecules \cite{Molecules}, in this way realizing a quantum gas with a particularly large dipolar interaction \cite{Aymar2005}. Finally, a two-species mixture offers the additional possibility to tune interactions and to conveniently apply component-selective methods. The main result reported in this letter is the first production of such a quantum degenerate two-species Fermi-Fermi mixture opening the door to aforementioned unexplored types of quantum matter. This goal was attained by achieving efficient sympathetic cooling of fermionic $\Li$ and $\K$ by an evaporatively cooled bosonic $\Rb$ gas. Moreover, we have also realized the first triple quantum degenerate mixture (see fig.\ref{fig:Clouds}), and therefore will be able to compare quantum properties of Fermi-Fermi and Bose-Fermi mixtures directly.

The basic idea of our experimental strategy is to sympathetically cool the fermions by a large rubidium cloud. In this way, the atom numbers of the fermions are in principle not reduced by evaporation and the initial fermion clouds can be loaded with reduced experimental effort. However, the challenge is to combine the different constraints which the individual atomic species enforce on the set of trapping and cooling parameters. Especially, the lack of sub-Doppler cooling of lithium, as well as the small elastic scattering cross section between rubidium and lithium \cite{Silber2005}, and the large Rb-Li mass ratio are not favorable preconditions. Nevertheless, we show that under such conditions the presence of a third atomic species can lead to more efficient sympathetic cooling (``catalytic'' cooling). We demonstrate this by investigating the final stage of the sympathetic cooling process into quantum degeneracy, which is of particular interest for further studies with various mixtures.

\begin{figure}
 \centering
 \includegraphics[width=8cm]{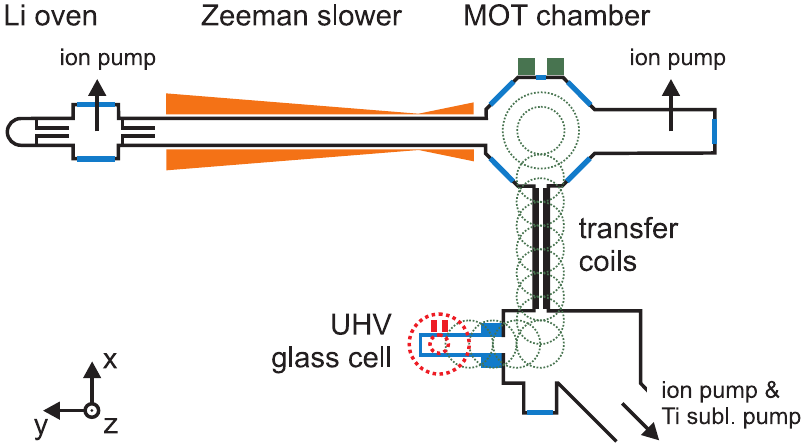}
 \caption{\label{fig:apparatus} Schematic of the apparatus in top view. The two fermionic and the bosonic species are magneto-optically trapped in a first vacuum chamber. They are then magnetically transferred into an UHV glass cell, where they are simultaneously cooled into quantum degeneracy.}
\end{figure}

The concept of the apparatus is an extension of the triple magneto-optical trap (MOT) setup previously described in \cite{Taglieber2006}. The experimental setup is shown in Fig.\,\ref{fig:apparatus}. Initially, cold clouds of all three species are magneto-optically trapped at the center of a common magnetic quadrupole field in a first chamber (MOT chamber). Lithium is loaded from a spin-flip Zeeman slower. $\Rb$ and $\K$ are loaded from the background vapor produced by dispensers. The home-built potassium dispensers were upgraded to enriched potassium with an abundance of 6\% in $\K$. The three atomic clouds are then transferred into a magnetic quadrupole trap in the MOT chamber. In order to attain a longer trap lifetime, the atoms are subsequently transferred into a second chamber, an ultra high vacuum (UHV) glass cell, with a residual pressure below \mbox{$10^{-11}$\,mbar}. This transfer is realized by driving a sequence of shifted quadrupole coils \cite{Greiner2001}. Evaporative and sympathetic cooling into the quantum degenerate regimes are then carried out in a quadrupole-Ioffe-configuration trap (QUIC) \cite{Esslinger1998}. At the center of this trap, the $\Rb$ radial (longitudinal) oscillation frequency is $\omega_\bot/2\pi=156.5(1)$\,Hz ($\omega_\|/2\pi=20.32(2)\,\mathrm{Hz}$) at a bias field of $B_0=3.2\,\mathrm{G}$. The trapping frequencies for $\Li$ and $\K$ are a factor $\sqrt{m_\mathrm{Rb}/m_\mathrm{Li}}\approx 3.81$ and $\sqrt{m_\mathrm{Rb}/m_\mathrm{K}}\approx 1.47$ higher.

The experimental cycle to produce the mixture of quantum degenerate gases starts with loading of the triple MOT for $15\,\mathrm{s}$ at an axial magnetic field gradient of $15\,\mathrm{G/cm}$. For efficient loading into the magnetic trap, it is necessary to obtain low temperatures and high densities for all three clouds simultaneously. However, standard laser cooling does not work equally well for all three species. We therefore apply a temporal dark MOT (dMOT) for rubidium and a compressed MOT (cMOT) for lithium. The cMOT/dMOT phase is initiated by a linear ramp of the magnetic field gradient to \mbox{28\,G/cm} within \mbox{25\,ms} at a reduced rubidium repumper intensity of approximately \mbox{1\%} of the saturation intensity. The ramp is followed by \mbox{5\,ms} at a large detuning of \mbox{$-72$\,MHz} and slightly decreased intensity for the rubidium trapping light. For the last \mbox{2\,ms} of this phase, the intensities of trapping and repumping light for lithium are reduced to approximately \mbox{3\%} of the saturation intensity and the detunings are reduced from \mbox{$-32$\,MHz} to \mbox{$-11$\,MHz} and \mbox{$-21$\,MHz}, respectively. During the whole cMOT/dMOT phase, the detunings and intensities of the potassium light are kept at the values used for MOT loading. This scheme increases the lithium density to $1.5\times 10^{10}\,\mathrm{cm}^{-3}$ \cite{ValueRemark} and decreases the temperature to $520\,\mu\mathrm{K}$, both improvements by more than a factor of two with respect to the MOT. The dMOT/cMOT phase is followed by \mbox{1.5\,ms} of optical molasses cooling for rubidium only, resulting in a temperature of \mbox{45\,$\mu$K} and a density of approximately $2\times 10^{11}\,\mathrm{cm}^{-3}$, improvements by factors of approximately 15 and 3 as compared to the MOT. 

For magnetic trapping, the atoms are prepared in the maximally polarized states $\textrm{Rb}\left|F=2,m_\mathrm{F}=2\right\rangle$, $\Kmp$, and $\Limp$ by means of optical pumping. This is the only magnetically trappable mixture that is stable against decay through spin-exchange collisions. The atomic clouds are then captured in a magnetic quadrupole field with a gradient of \mbox{140\,G/cm} along the quadrupole axis. This capture gradient does not preserve phase space density for the three species but is a good compromise between the different requirements. Subsequently, the quadrupole potential with the trapped clouds is moved in \mbox{2.5\,s} over a distance of \mbox{39\,cm} around a $90^\circ$ corner into the glass cell. After the transfer, the atomic clouds are compressed by increasing the magnetic field gradient to \mbox{300\,G/cm} within \mbox{2\,s}. The quadrupole potential is then transformed into the QUIC potential within \mbox{3.3\,s}. A temporary overcompensation of the trap bias field was added to the original transformation mechanism described in \cite{Esslinger1998}. This increases the calculated effective trap depth during the transfer from \mbox{0.9\,mK} to \mbox{2.1\,mK} and prevents significant loss of atoms from the hot lithium cloud at the wall of the glass cell. 

Subsequently, the evaporative and sympathetic cooling process is started. Species-selective evaporative cooling of rubidium atoms in the $\Rbmp$ state is achieved by driving the transition to the $\RbOneOne$ hyperfine state using microwave (MW) radiation. Undesired populations in the two other magnetically trappable rubidium states $\RbTwoOne$ and $\RbOneMOne$ have to be removed by an additional MW cleaning signal as a necessary precondition to reach BEC. The evaporation sequence is initiated by a \mbox{1.2\,s} long sweep from \mbox{43\,MHz} below the $\RbOneMOne\rightarrow\RbTwoZero$ transition to \mbox{2\,MHz} above the $\RbTwoOne\rightarrow\RbOneZero$ transition at the trap bottom. This sweep removes atoms in the $\RbTwoOne$ and $\RbOneMOne$ states from the trap. Subsequently, the evaporation ramp is started. We observe that the $\RbTwoOne$ state is constantly repopulated during the evaporation process. It therefore has to be cleaned permanently. This is implemented by ramping the cleaning signal linearly from \mbox{150\,kHz} above to close to the $\RbTwoOne\rightarrow\RbOneZero$ transition frequency at the trap bottom. Alternative cleaning schemes with repeated ramps gave inferior results.

The duration of the sympathetic cooling process depends on the inter-species thermalization rate. The $\K$-$\Rb$ inter-species scattering cross section is relatively large \cite{Ferlaino2006} and we successfully cooled $\K$ into quantum degeneracy using a \mbox{29\,s} long MW-evaporation ramp for rubidium. However, sympathetic cooling of $\Li$ by $\Rb$ is more challenging because thermalization is considerably slower due to the small $\Li$-$\Rb$ inter-species scattering cross section \cite{Silber2005}, which is roughly two orders of magnitude smaller than the $\K$-$\Rb$ cross section (in the low-temperature limit). In addition, the larger mass difference results in both a lower energy transfer per elastic collision, which is only partially compensated by a higher mean thermal relative velocity, and a reduced density overlap of the two clouds. For sympathetic cooling of lithium, we therefore use an evaporation sequence that is stretched in time with respect to a sequence optimized for the production of rubidium BEC. Experimentally, we found gradually increased stretching towards the end of the evaporation ramp to be more beneficial than stretching at the beginning. For the work presented here, an optimized ramp with a total duration of \mbox{63\,s} was used. However, we found that cold lithium clouds with significant atom numbers can only be achieved, if lithium atoms that are left behind in the sympathetic cooling process (i.e. which have an energy much higher than the average energy of the rubidium atoms) are constantly removed from the trap. This is realized by applying radio frequency radiation on the hyperfine transition of the lithium ground state. This Li hf-knife is ramped linearly in \mbox{63\,s} from a cut energy of \mbox{3\,mK$\times\kB$} to \mbox{4\,$\mu$K$\times\kB$} during the evaporation ramp. Experimentally, we find that lower initial cut energies result in lower final lithium atom numbers. This is a strong indication that not only atoms in the low-energy tail of the initial lithium cloud are sympa\-the\-tically cooled but also high-energy lithium atoms with low angular momentum.

At the end of the experimental sequence absorption images of the clouds are taken along the z-axis. The deduced optical column density is then projected along the x- or the y-axis to improve the signal to noise ratio. The obtained optical line density is fitted with a fully physical model function. In the case of fermions, an integrated Fermi-Dirac density distribution is used, from which the degeneracy parameter $T/T_F$ and the atom number $N$ can be directly calculated. We conservatively estimate the systematic uncertainty of the atom numbers to be below 50\% and of the temperatures to be lower than $\pm0.1\,T_\mathrm{F}$ in the temperature range of 0.2--0.5$\,T_\mathrm{F}$. In the case of a partially condensed rubidium cloud, we use an appropriate, integrated two-component fit function consisting of the sum of a Bose density distribution and a Thomas-Fermi density distribution. The temperature and the atom number in the thermal fraction are directly calculated from the best fit parameters. The atom number in the condensed fraction is deduced from the fitted Thomas-Fermi radius and the known rubidium scattering length. 

\begin{figure*}[!t]
 \centering
 \includegraphics[width=15cm]{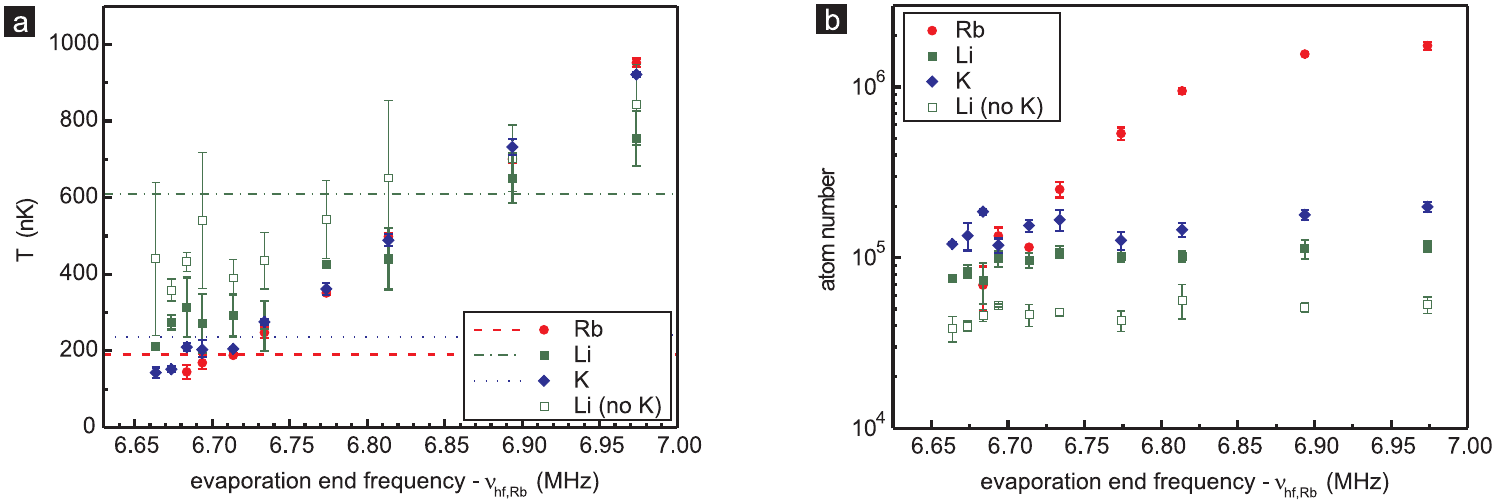}
 \caption{\label{fig:TandN} Temperatures (a) and atom numbers (b) during the very last part of simultaneous trapping and cooling of three species (filled symbols). $\Rb$ is evaporatively cooled, $\Li$ and $\K$ are sympathetically cooled by thermal contact with the rubidium cloud. Corresponding data for lithium in a two-species $\Li$-$\Rb$-mixture are also shown (empty rectangles). 
The horizontal lines in (a) indicate the critical temperature $T_\mathrm{c}$ for rubidium and half the Fermi temperature $T_\mathrm{F}$ for the fermions, respectively. For our trap parameters and typical atom numbers of $1.5 \times 10^5$ for rubidium and $10^5$ for the fermionic species, the critical temperature is $T_\mathrm{c}^\mathrm{Rb}=190\,\mathrm{nK}$ and the Fermi temperatures are \mbox{$T_\mathrm{F}^\mathrm{Li}=1.2\,\mu\mathrm{K}$} and \mbox{$T_\mathrm{F}^\mathrm{K}=470\,\mathrm{nK}$}. The error bars represent the r.m.s.-deviation for at least three consecutive repetitions.}
\end{figure*}
The temperatures and atom numbers during the last part of the evaporation ramp are shown in \mbox{Fig.\,\ref{fig:TandN}}. The simultaneously trapped mixture has been repeatedly produced for different end frequencies of the MW-evaporation ramp and for separate imaging of the three species. The data show that simultaneous quantum degeneracy for the three-species mixture $\Li$-$\K$-$\Rb$ was achieved. For partially condensed rubidium clouds the atom numbers of the thermal fraction are given. For the very lowest two evaporation end frequencies, rubidium clouds were absent or too small in atom number to be fittable. Moreover, the temperature data show that rubidium and potassium are very well thermalized throughout the temperature range observed. However, for the reasons mentioned above, the lithium cloud is not fully thermalized with the rubidium atoms even for the slow evaporation ramp in use. We also investigated sympathetic cooling of a two-species $\Li$-$\Rb$ mixture by omitting loading of potassium. The comparison of the lithium temperatures in the two situations clearly shows that lithium can be cooled more efficiently, when also potassium is present in the trap during the cooling process. This is further supported by the observed atom numbers: In the full $\Li$-$\K$-$\Rb$ three-species mixture, the atom numbers for both fermionic species are nearly constant at approximately $10^5$ in the observed part of the cooling process. However, the lithium atom number is a factor of 2 smaller, if only the two-species mixture $\Li$-$\Rb$ is used. This indicates that the $\Li$-$\K$ thermalization rate is comparable to or even larger than the $\Li$-$\Rb$ thermalization rate, at least at some temperature during the evaporation ramp. Since the potassium atom number is constant and the rubidium and potassium clouds are well thermalized, the energy of the lithium cloud is eventually transferred to the rubidium cloud. In this sense, potassium acts as a catalytic cooling agent for lithium.

A typical example of a quantum degenerate three-species mixture is shown in \mbox{Fig.\,\ref{fig:Clouds}}. For potassium, fitting with an appropriate Fermi-Dirac density distribution function yields an atom number of \mbox{$N_\mathrm{K}=1.3\times 10^5$} and a temperature of \mbox{$T_\mathrm{K}=184\,\mathrm{nK}=0.35\,T_\mathrm{F}$}. The corresponding numbers for lithium are \mbox{$N_\mathrm{Li}=0.9\times 10^5$} and \mbox{$T_\mathrm{Li}=313\,\mathrm{nK}=0.27\,T_\mathrm{F}$}. For rubidium, fitting with an appropriate two-component density distribution gives \mbox{$N_\mathrm{thermal, Rb}=1.5\times 10^5$} and \mbox{$T_\mathrm{Rb}=189\,\mathrm{nK}$} for the thermal fraction and an atom number of \mbox{$N_{0,\mathrm{Rb}}=4.5\times 10^4$} in the condensed fraction. State pureness for all three species was verified using a Stern-Gerlach method.

The experimental sequence described above was optimized for simultaneous quantum degeneracy of all three species. For this purpose, the heat load on rubidium had to be reduced by loading less lithium atoms into the MOT. If a quantum degenerate mixture of the fermionic species $\Li$ and $\K$ only is desired, however, the lithium MOT is loaded to maximum atom number and rubidium is completely evaporated from the trap. This procedure then results in a mixture of \mbox{$1.8\times 10^5$} lithium atoms at a temperature ratio of \mbox{$T/T_\mathrm{F}=0.34$} and \mbox{$1.8\times 10^5$} potassium atoms at \mbox{$T/T_\mathrm{F}=0.40$}. This mixture is an optimal starting point for Fermi-Fermi experiments.

In conclusion, we have demonstrated the first quantum degenerate Fermi-Fermi mixture of two different atomic species. In addition, a triple quantum degenerate Bose-Fermi-Fermi mixture was realized. Sympathetic cooling can yield sufficiently high degeneracy parameters and atom numbers in spite of the very different initial temperatures and masses of the three species. We can now use this experimental platform to study the Fermi-Fermi as well as Bose-Fermi mixtures. Projects that seem especially interesting here are the investigation of the effect of the mass difference on the superfluid phase, the creation of heteronuclear molecules, and the exploitation of species-selective trapping potentials.

We would like to thank C.\,Ei\-gen\-wil\-lig, S.\,Fray, F.\,Hen\-kel, and W.\,Wieser for experimental assistance during the setup of the apparatus. This research was supported by the DFG cluster of excellence Munich Centre for Advanced Photonics (www.munich-photonics.de).


\begin{thebibliography}{10}

\bibitem{BEC}
M.H.~Anderson \etal, Science {\bf 269}, 198 (1995); K.B.~Davis \etal, Phys. Rev. Lett. {\bf 75}, 3969 (1995).

\bibitem{Greiner2002}
M.~Greiner \etal, Nature {\bf 415}, 39 (2002).

\bibitem{Fermions}
B.~DeMarco and D.S.~Jin, Science {\bf 285}, 1703 (1999); A.G.~Truscott \etal, Science {\bf 291}, 2570 (2001).

\bibitem{BecBcs}
M.~Holland, S.J.J.M.F.~Kokkelmans, M.L.~Chiofalo, and R.~Walser, Phys. Rev. Lett. {\bf 87}, 120406 (2001); C.~Chin \etal, Science {\bf 305}, 1128 (2004); C.A.~Regal, M.~Greiner, D.S.~Jin, Phys. Rev. Lett. {\bf 92}, 040403 (2004); M.W.~Zwierlein \etal, Nature {\bf 435}, 1047 (2005); Q.~Chen \etal, Phys. Rep. {\bf 412}, 1 (2005); R.~Haussmann, W.~Rantner, S.~Cerrito, and W.~Zwerger, Phys. Rev. A {\bf 75}, 023610 (2007); S.~Giorgini, L.P.~Pitaevskii, S.~Stringari, preprint at $\left\langle \mathrm{http://arXiv.org/abs/0706.3360v1}\right\rangle$ (2007).

\bibitem{FermiFermi}
D.S.~Petrov, C.~Salomon, G.V.~Shlyapnikov, Phys. Rev. Lett. {\bf 93}, 090404 (2004); C.A.~Regal, M.~Greiner, D.S.~Jin, Phys. Rev. Lett. {\bf 92}, 083201 (2004); K.~Dieckmann \etal, Phys. Rev. Lett. {\bf 89}, 203201 (2002).

\bibitem{ManyBody}
R.~Casalbuoni, G.~Nardulli, Rev. Mod. Phys. {\bf 76}, 263 (2004); T.~Mizushima, M.~Ichioka, K.~Machida, J. Phys. Chem. Sol. {\bf 66}, 1359 (2005); S.-T.~Wu, C.-H.~Pao, S.-K.~Yip, Phys. Rev. B {\bf 74}, 224504 (2006); H.~Hu, X.-J.~Liu, P.D.~Drummond, Phys. Rev. Lett. {\bf 98}, 070403 (2007).

\bibitem{Petrov2007}
D.S.~Petrov \etal, preprint at $\left\langle \mathrm{http://arXiv.org/abs/0706.2855v1}\right\rangle$ (2007).

\bibitem{QCD}
F.~Wilczek, Nature Physics {\bf 3}, 375 (2007), News and Views; \'A.~Rapp, G.~Zar\'and, C.~Honerkamp, and W.~Hofstetter, Phys. Rev. Lett. {\bf 98}, 160405 (2007).

\bibitem{Molecules}
S.~Azizi, M.~Aymar, O.~Dulieu, Eur. Phys. J. D {\bf 31}, 195 (2004);
J.~Doyle \etal, Eur. Phys. J. D {\bf 31}, 149 (2004).

\bibitem{Aymar2005}
M.~Aymar, O.~Dulieu, J. Chem. Phys. {\bf 122}, 204302 (2005).

\bibitem{Silber2005}
C.~Silber \etal, Phys. Rev. Lett. {\bf 95}, 170408 (2005).

\bibitem{Taglieber2006}
M.~Taglieber \etal, Phys. Rev. A {\bf 73}, 011402(R) (2006).

\bibitem{Greiner2001}
M.~Greiner, I.~Bloch, T.W.~H\"ansch, and T.~Esslinger, Phys. Rev. A {\bf 63}, 031401(R) (2001).

\bibitem{Esslinger1998}
T.~Esslinger, I.~Bloch, and T.W.~H\"ansch, Phys. Rev. A {\bf 58}, R2664 (1998).

\bibitem{ValueRemark}
The densities and temperatures listed in this paragraph are typical for respective single species operation. In three species operation, the MOT atom numbers are lower by less than a factor of two. 

\bibitem{Ferlaino2006}
F.~Ferlaino \etal, Phys. Rev. A {\bf 73}, 040702(R) (2006) and ibid. {\bf 74}, 039903(E) (2006).

\end{thebibliography}
\end{document}